\documentclass{raa}            
\usepackage{longtable}

\usepackage{graphicx,times}             

\begin{document}

   \title{The Fifth Edition of the General Catalogue of Variable Stars: Experiences in the Constellation of Centaurus
   }

   \volnopage{Vol.0 (200x) No.0, 000--000}      
   \setcounter{page}{1}          

   \author{N. N. Samus
      \inst{1}
   \and E. N. Pastukhova
      \inst{1}
   \and O. V. Durlevich
      \inst{2}
   \and E. V. Kazarovets
      \inst{1}
   }

   \institute{Institute of Astronomy, Russian Academy of Sciences, 48, Pyatnitskaya Str., Moscow, 119017, Russia; {\it samus@sai.msu.ru }\\
        \and
             Sternberg Astronomical Institute, Lomonosov Moscow University, 13, University Ave., Moscow, 119234 Russia \\
   }

   \date{Received~~2018 month day; accepted~~2018~~month day}

\abstract{We have recently announced that the General Catalogue of
Variable Stars enters the stage of its fifth,  purely electronic
edition (GCVS 5.1). Recently, we have incorporated a new,
completely revised version for 1408 variable stars in the
constellation of Centaurus into the GCVS 5.1. Working on this
revision, we used current possibilities of data mining, suggested
new variability types for many variable stars, found new light
elements for a large number of periodic variables. This paper
describes the work fulfilled during the preparation of the GCVS
5.1 version for Centaurus, discusses in detail a number of cases
most significant astrophysically.  }

   \authorrunning{N. N. Samus et al.}            
   \titlerunning{The Fifth Edition of the GCVS: Experiences in Centaurus}  

   \maketitle

%
%
\section{Introduction}           
\label{sect:intro}

We briefly discussed the history of variable-star catalogues
published in the USSR and Russia in Samus' et al. (2017). After
the World War~II, the catalogues are being compiled in our country
on behalf of the International Astronomical Union. The General
Catalogue of Variable Stars (GCVS) has the purpose to include
reliable variable stars: as a rule, with variability types, at
least tentative ones, known. Stars with doubtful variability or
with unknown variability types should be included into Catalogues
of Suspected Variable Stars. Since 1948, four major editions of
the GCVS, in the book form, were published. The fourth edition
(Kholopov et al., 1985--1995) consisted of five volumes and
contained 28435 variable stars in its main part (Volumes I--IV)
plus almost 12000 variable stars in external galaxies and
extragalactic supernovae in its Vol.~V. Catalogues of Suspected
Variable Stars, published by now (Kukarkin et al., 1982;
Kazarovets et al., 1998), contain more than 26000 stars; about
4500 of them have already got their GCVS names by now.

After the completion of the 4th edition, the GCVS team identified
almost all GCVS stars with astrometric catalogues or, in a number
of cases when such identification was not possible, measured
accurate coordinates using available images. Identification was
found completely impossible (only rough coordinates published;
lacking charts) for about 240 stars, a very small fraction. For a
vast majority of GCVS objects, it is now possible to find them
using solely the coordinates presented in the GCVS.

Samus' et al. (2017) announced the current electronic version of
the GCVS (http://www.sai.msu.su/gcvs/gcvs/) to be the version GCVS
5.1; Samus' et al. (2017) is now considered the standard reference
to this GCVS version. As of June 2017 (after publication of the
most recent Name-list, see below), the version contains 51853
variable stars (not counting entries corresponding to non-existing
stars or to objects that, by mistake, obtained two or even three
different GCVS names, mainly because of published wrong
coordinates).

Stars enter the GCVS through the so-called Name-lists of variable
stars. One of the main current problems of the GCVS is that the
Name-lists Nos. 67--77 that had appeared after the 4th GCVS
edition contained only coordinates, variation magnitude ranges,
and variability types but did not present such important
information as light elements (for periodic variable stars) or
spectral types. The subsequent Name-lists Nos. 78--81 (more than
13300 variable stars) are mini-catalogues of variable stars,
containing all the above-mentioned kinds of data.

Our further work will eventually result in the GCVS 5.2 version.
This work consists in:

--  preparing new Name-lists providing, for stars added to the
GCVS, all information that should be contained in the GCVS;

-- filling the information gap (due to incompleteness of the
Name-lists Nos. 67-77) with all relevant data;

--   updating the information provided for the stars of the 4th
GCVS edition and for stars from relatively recent Name-lists
taking into account new publications and processing photometric
data available through data mining.

\noindent The last two tasks are being gradually performed in the
alphabet order of Latin names of the constellations\footnote{Note
that, due to tradition, GCVS names consist of the variable star's
name proper and the Genitive of the constellation name, like RR
Lyrae. For the description of this system, see
http://www.sai.msu.su/gcvs/gcvs/gcvs5/htm/.}. By 2017, this work
was completed for 18 constellations (Andromeda--Cassiopeia). This
paper presents the results for the 19th constellation, Centaurus.

\section{GCVS Revision in Centaurus}

As of February 2018, the GCVS 5.1 contains 1408 variable stars in
the constellation of Centaurus (actually, the main GCVS table for
Centaurus consists of 1412 lines, but four entries are for star
names no longer recommended for use: at some time, these stars
erroneously obtained their second GCVS names).  Note that there
were only 833 Centaurus variable stars in the 4th GCVS edition.
The Name-lists Nos. 67-77 presented limited information for 216
variable stars. More detailed information for 359 variable stars
can be found in Name-lists Nos. 78--81, but new sources of data
provide possibilities of improving GCVS information even for some
of  them.

To improve the GCVS, we used variable-star studies published in
many papers from the recent astronomical literature. Whenever
possible, we checked the published data ourselves by means of
photometric data mining. The main sources of data mining we used
in this study in order to improve the GCVS data in Centaurus were
the ASAS-3 photometric survey (Pojmanski, 1997) and the Catalina
Sky Surveys (Drake et al., 2009). The ASAS-3 survey provides
$V$-band photometry for stars between approximately magnitudes 7
and 14.5; it covers the whole sky to the south of declination
+30$^\circ$. The survey is bases on observations with very small
telescopes, and thus the angular resolution in crowded fields is
poor. The Catalina Surveys are based on observations with CCDs
sensitive to red light, but the magnitudes were calibrated using
$V$-band standard stars. The Catalina magnitudes are sometimes
designated $CV$. The telescopes used in this survey are
moderate-sized, providing much better angular resolution. The
survey covers northern and southern sky, but with the sky poles
and the Milky Way strip excluded. In Centaurus, it can be used
only in the parts of the constellation most distant from the Milky
Way. Bright stars are overexposed, the best results can be
achieved for stars in the $13^m-17^m$ magnitude range. The
observation times provided by the ASAS-3 survey are heliocentric,
but the Catalina Surveys present geocentric times, which we
converted to heliocentric times for short-period variable stars.

\section{Results}

Accurate coordinates of variable stars in Centaurus were presented
by Samus' et al. (2002). When working on the GCVS 5.1 version for
Centaurus, we revised identifications or improved coordinates for
35 variable stars.

In the 4th GCVS edition, among the 833 variable stars in
Centaurus, only 20 had first references  to the research of the
GCVS authors. For the same 833 stars, we now have 549 references
(66\%) to our research, performed mainly by data mining. In total
for the 1408 Centaurus stars, the number of references to the
research of the GCVS team is currently 793 (56\%).

Among the variables AF Cen--IT Cen (180 stars in total), 135 stars
had references to Hoffleit (1930) in our 4th edition. The GCVS 5.1
version for Centaurus currently has only 38 references to this
paper (for stars with no possibilities to improve the data using
data mining). This is the first paper ever published by Dorrit
Hoffleit (1907--2007). In her autobiography published by the AAVSO
(Hoffleit, 2002), she remembers that, when preparing her first
scientific publications, she even did not know about the
possibility of spurious periods. Our analysis shows, as the most
frequent mistake, that red semiregular or irregular variable stars
were erroneously believed to be stars with rapid brightness
variations (the author did not use color information to check her
results). The GCVS 5.1 classification for some of these stars is
confirmed by spectral types; in other cases, infrared color
indices are known from the 2MASS catalog. Table~1 summarizes the
most important changes in the GCVS 5.1 for Hoffleit's stars.
Columns 2 and~3 reproduce GCVS~4 type and period, taken from
Hoffleit (1930), and the following columns are GCVS 5.1 data,
mainly determined by us. ``VSX'' means information from the
International Variable Star Index (Watson et al., 2007).

The light curve of HQ Cen, the eclipsing (EB) star with a rather
long period, is presented in Fig.~1. This star was sufficiently
correctly solved in the ASAS-3 catalog (Pojmanski, 1997).

Among the variables V444 Cen--V490 Cen (46 stars in total: the
name V467 Cen is no longer recommended for use, the current name
of the star is V746 Cen), 41 variable stars had references to
Huruhata (1940), again a study based on plates of the Harvard
stacks. Only three stars remain with references to Huruhata (1940)
in the GCVS 5.1. We were able to solve many stars using Catalina
observations. In most cases, the changes were just an improvement
of the period, but there were also more serious corrections
(spurious periods in Huruhata, 1940). The most important changes
in the GCVS 5.1 for Huruhata's stars are collected in Table~2.
Columns 2 and 3 reproduce GCVS~4 type and period, taken from
Huruhata (1940), and the following columns are GCVS 5.1 data,
mainly determined by us.

Among the variables V500 Cen--V569 Cen (70 stars), all but one
have GCVS~4 references to McLeod and Swope (1941). This study in
Harvard plate stacks resulted in much better periods than those
determined in Huruhata (1940). Nevertheless, the revision of these
results in the GCVS 5.1 was also significant (Table~3). Columns~2
and~3 reproduce GCVS~4 type and period, taken from McLeod and
Swope (1941), and the following columns are GCVS 5.1 data, mainly
determined by us.

\begin{figure}
   \centering
   \includegraphics[width=\textwidth, angle=0]{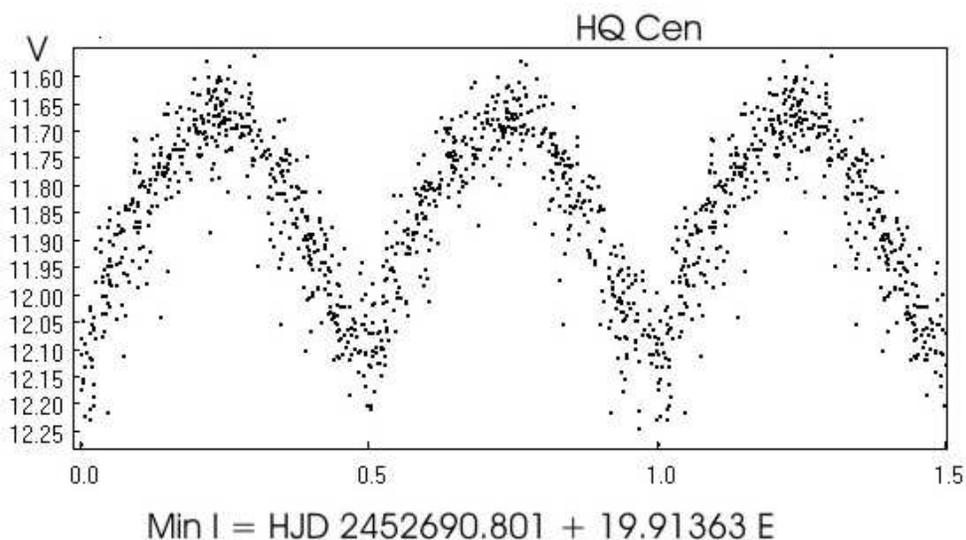}
   \caption{The light curve of HQ Cen, from ASAS-3 observations.}
   \label{Fig1}
   \end{figure}

Data for eight variable stars (V647 Cen--V654 Cen) in the 4th GCVS
edition were based on Ninger-Kosybowa (1949). All these stars have
now been revised by the GCVS 5.1 compilers (Table~4). Especially
impressive is the case of V653 and V654~Cen that are definitely
W~UMa eclipsing variables rather than rapid irregular stars
(expression used in the original paper by Ninger-Kosybowa, 1949:
``Probably an irregular variable star with short variations of
brightness''). The light curves for these two stars are presented
in Figs.~2 and~3. Note the O'Connell effect (different heights of
maxima) for both variables, especially for V654 Cen.

\begin{figure}
   \centering
   \includegraphics[width=\textwidth, angle=0]{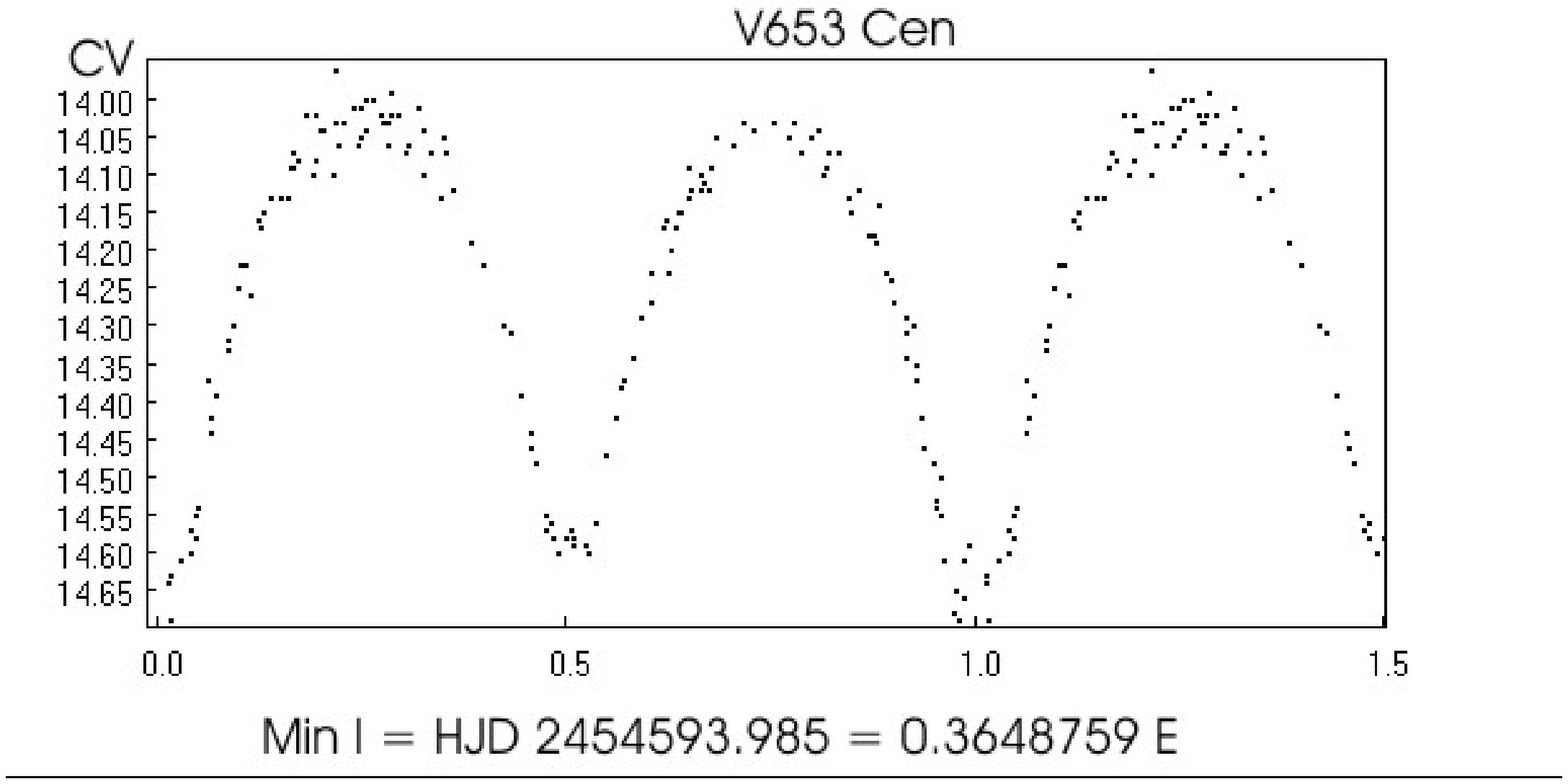}
   \caption{The light curve of V653 Cen, from Catalina observations.}
   \label{Fig1}
   \end{figure}

\begin{figure}
   \centering
   \includegraphics[width=\textwidth, angle=0]{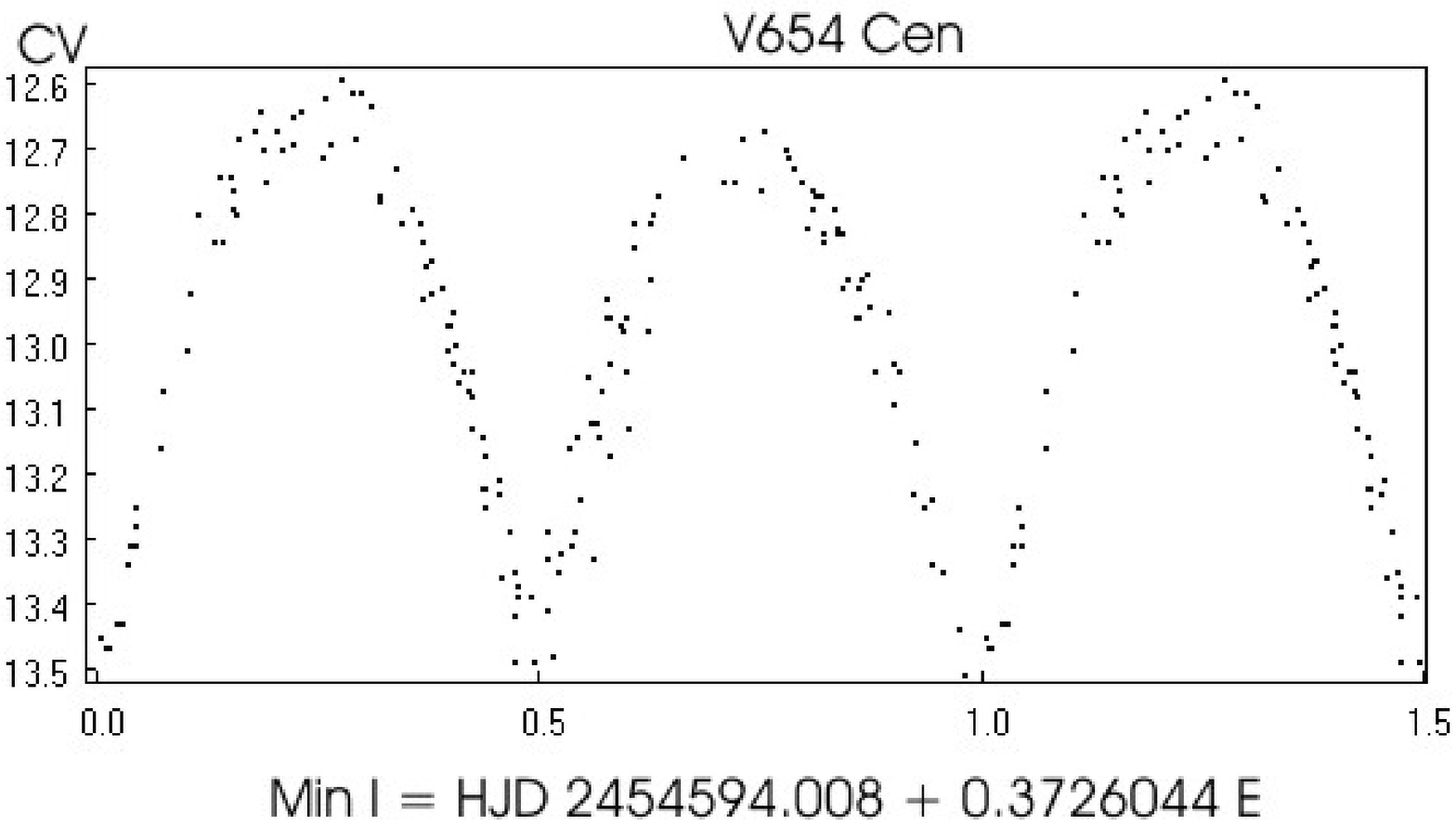}
   \caption{The light curve of V654 Cen, from Catalina observations.}
   \label{Fig1}
   \end{figure}

The famous German variable-star researcher Cuno Hoffmeister
(1892--1968), who made a large contribution also into studies of
southern variable stars (based on several observing trips to
Africa), did not work much in Centaurus. Twenty-nine variable
stars in the 4th GCVS edition (V714, V715, V718--V736 Cen,
V738--V741 Cen, V792--V794 Cen, V797 Cen) had references to
Hoffmeister (1963). In this paper, the author suggested
variability types on the base of very limited numbers of
photographic brightness estimates, usually not deriving any light
elements. Among the stars listed above, he gave the type ``RR''
(without subtype) for 22 variables (one of them uncertain); five
stars were announced as eclipsing variables and two stars, as slow
irregulars (only one of them definitely red). No light elements
were presented for any of the 29 stars. We have studied all these
stars using Catalina data (ASAS-3 data for V794 and V797 Cen),
Classification was confirmed for all the 22 RR Lyraes and for all
announced eclipsing stars, with the exception of V797 Cen. We
determined light elements and subtypes for RR Lyraes (20 of them
are RRAB variables), light elements for eclipsing stars. V792 Cen
(type L from Hoffmeister) is actually an EW eclipsing star; V794
Cen is an SRA variable (instead of LB), and V797 Cen is not an
eclipsing star but an SRB variable.

Here we briefly presented only several big blocks corresponding to
the transition between the 4th GCVS edition and GCVS 5.1 in
Centaurus. Cases of contribution from the GCVS authors to
information on individual stars are also numerous.

V667 Cen is an RR Lyrae variable discovered by Shapley et al.
(1954). They did not publish a finding chart, and their published
coordinates were not accurate enough, so a search in a rather wide
field was required, which consisted of checking stars in the
appropriate magnitude range for variations in photometric surveys.
Our attempts to find the variable in 2000--2004, supported by A.
Paschke (private communication, 2004), led to a wrong
identification. Luckily, when revising the GCVS in 2017 and
feeling unsure about the identification of V667 Cen, we repeated
the search using data from Catalina surveys and were able to
identify the variable correctly, with GSC 7303--00247. The period
of this RRAB star, determined by us from Catalina data, is
0.6637930 days, while Shapley et al. (1954) suggested a different
one, 0.60135 days.

V786 Cen entered the GCVS as a possible eclipsing variable
discovered by Ebisch (1973), with no light elements. As of
February 2018, the well-known Internet resource ``O--C Gateway''
suggested two possible periods for the star, 0.855098 and 1.19741
days. From Catalina data, we find both these periods to be wrong,
the correct one for this Algol (EA) star being 5.9864070 days.

V805 Cen was discovered by Strohmeier (1967) who had not
determined its variability type. W\"alder (1975) correctly
attributed the star to eclipsing variables, but her period,
2.211155 days, is wrong. The ``O--C Gateway'' site suggests two
solutions, that by W\"alder (1975) and that with a new period,
3.3166 days. This period is nearly correct but not accurate
enough. Our new light elements from ASAS-3 data are: $${\rm Min} =
2453560.652 + 3^d.3167193\times E.$$ The most interesting feature
of the light curve with this period is that the secondary minimum
is observed at phase 0.547. Such eccentric eclipsing stars are
important from the point of view of possible apsidal motion.

Several astrophysically important corrections for Centaurus stars
contained in the 4th GCVS edition were suggested by other authors.
Thus, the compilers of the fourth edition decided, on the base of
the discovery paper (Elvius, 1975), that V803 Cen belonged to the
rare RCB type. It is now clear that here we deal with a
cataclysmic, Nova-like variable (cf., for example, Patterson et
al., 2000). V819 Cen was believed to be a pulsating variable,
probably an RR Lyrae star, with P = 0.6755 days (Moffat, 1977).
The variable is now known to be a silicon Ap star (e.g., Maitzen
and Vogt, 1983), so we consider it an ACV variable star. Paschke
and Poretti (2006) determined its period to be 2.078588 days from
ASAS-3 data.

As noted above, Name-lists Nos. 67--77 presented only limited
information compared to that standard for the GCVS. In Centaurus,
these Name-lists contained 219 objects. GCVS 5.1 now gives
complete information for all of them, with references to GCVS
compilers' original work in 82 cases (37\%).

Name-lists Nos. 78--81 were prepared in the complete GCVS format.
Nevertheless, when preparing the GCVS 5.1 version for Centaurus,
we revised all these stars, introducing, whenever necessary, new
light elements and other corrections. For these 357 stars, there
are now 151 references to our work (42\%). Additionally, 27 stars
have references to the papers we published when preparing these
Name-lists (Kazarovets et al., 2005; Kazarovets and Pastukhova,
2007, 2008ab, 2009; Pastukhova, 2007).

\section{Concluding remarks}

The work on the GCVS 5.1 version is being continued. The GCVS team
is currently working on the next, 82nd Name-list of variable
stars, expected to contain several thousand variable stars of the
whole sky. We are finishing a complete revision of GCVS
information in the next constellation, Cepheus. Our nearest plans
include incorporating sufficiently well-studied variable stars in
globular clusters into the GCVS system (note that, for reasons of
tradition, the GCVS, intended to be a catalog of all {\it
galactic} variable stars, so far contained variables in open
clusters but, as a rule, not in globular clusters).

The number of photometrically variable stars, with variations
detectable using modern ground-based techniques, is very large.
Discoveries from space-borne observatories are able to detect even
smaller brightness variations. Note that astronomical tradition
never introduced any lower limit for the amplitude a star should
have to be called variable, making almost any star a potential
variable. In not so distant future, it will become not practical
to continue traditional naming of variable stars. We expect (Samus
and Antipin, 2015) that future large general-purpose star
catalogues will contain sections with characteristics of
brightness variations for sufficiently well-studied stars.
Meanwhile, the astronomical scientific community still expresses
interest in having traditional GCVS names for their new
discoveries. As long as it remains technically possible, we will
continue compiling new Name-lists and updating GCVS information.

\begin{acknowledgements}

We would like to thank Dr. A.V. Khruslov for his active assistance
during several recent years. Our work on variable-star catalogues
is supported, in part, by the sub-programme ``Astrophysical
Objects as Space Laboratories'' in the Programme P-28 of the
Presidium of Russian Academy of Sciences.

\end{acknowledgements}

\newpage

\begin{table}
\begin{center}
\caption[]{Variable stars from Hoffleit (1930) in GCVS 5.1.}\label{Tab:hoffl}

 \begin{tabular}{llllll}
  \hline\noalign{\smallskip}
GCVS name &  Type  & Period (GCVS 4), & Type & Period (GCVS 5.1), &Spectral type \\
          & (GCVS 4)&days             &(GCVS 5.1)&days&(GCVS 5.1)\\
\hline\noalign{\smallskip}
BQ Cen & S: & -- &   EW & 0.5468423 &  --\\
BU Cen & RV:& 85.5 & RVA & 170.4 & --\\
CP Cen & S: & --  & SRB & 147.7 &  M4\\
CR Cen & M  & 180.5& M &347& M8\\
CV Cen & SR & 150 &M  & 320: &   M9\\
CW Cen & S: & -- & SRB& 122 &--\\
CZ Cen & SR &--  & SRB& 132.5&--\\
DD Cen & S: &-- &SRB&44.7&--\\
DE Cen & E  & $1.0/N$ & EA & 0.59004 (VSX)&--\\
DF Cen & L  &--&SRD& 56.2&K4--K5\\
DH Cen & E  & $2/N$ &EA & 0.870765&--\\
DQ Cen & S: &--&SRB &190.5&   M6\\
DS Cen & RR &--& EW:&0.47512:&--\\
DT Cen &--&--&SRB:&36:&--\\
DV Cen & E&--&EA &1.205901&--\\
DX Cen &--&--& M:& 296 &M\\
EL Cen & E &--& EA:&21.8674:&--\\
EN Cen & E & $3/N$& EA&1.775753&--\\
EQ Cen &RR:&--& SRB &99.2 & M6\\
ES Cen & M & 173.6& M& 355.2&--\\
EU Cen & S:&--& M:&452&--\\
EV Cen & S:&--& LB&--&--\\
FF Cen &--&--&SRB&71.1&--\\
FI Cen & S:&--&SRB&195&--\\
FL Cen &RR:&--&LB:&--&--\\
FN Cen &S: &--&SRB&1059&--\\
FO Cen &S: &--&LB:&--&--\\
FS Cen &RR &--&RRAB:&0.50860:&--\\
FT Cen &RR &--&LB:&--&--\\
FU Cen &S: &--&LB:&--&--\\
FV Cen &L  &--& M & 249&--\\
FY Cen &-- &--&LB:& -- &--\\
FZ Cen &S:&--&LB:&--&--\\
GI Cen &S:&--&RRAB&0.635 &--\\
       &  &  &    &(Zorotovic et al. 2010)\\
GK Cen &RRAB&0.6599&CWB:&1.949773&--\\
GM Cen & S: &--&SRB&306&--\\
GN Cen & S: &--&SRB:&98:&--\\
GO Cen & RR:&--&LB:&--&--\\
GQ Cen & S: &--&SRB:&56:&--\\
GT Cen & -- &--&LB:&--&--\\
GU Cen & RR:&--&LB:&--&--\\
GV Cen & L  &--&SR:&--&M6-M7\\
GZ Cen & S: &--&M: &345:&--\\
HH Cen & S:&--& SRB& 61.1:&--\\
  \noalign{\smallskip}\hline
\end{tabular}
\end{center}
\end{table}

\addtocounter{table}{-1}

\begin{table}
\begin{center}
\caption[]{Continued}\label{Tab:hoffl}

 \begin{tabular}{llllll}
  \hline\noalign{\smallskip}
GCVS name &  Type  & Period (GCVS 4), & Type & Period (GCVS 5.1), &Spectral type \\
          & (GCVS 4)&days             &(GCVS 5.1)&days&(GCVS 5.1)\\
\hline\noalign{\smallskip}
HK Cen & E &$7/N$&EA&2.421757&--\\
HN Cen & RR&--&SRB&79&--\\
HQ Cen & S:&--&EB&19.91363&--\\
HS Cen &L: &--& SRB: &327:&--\\
HU Cen &S: &--& SRB: &72.6&--\\
HX Cen &E: &--& SRB: &81.9&--\\
IL Cen &S: &--& SRB  &422 &--\\
IN Cen &E  &--& RVA  &75.8196&--\\
IP Cen &M  &188.5& M &377&--\\
IT Cen &RR &=-& SRB &110&--\\
  \noalign{\smallskip}\hline
\end{tabular}
\end{center}
\end{table}

\begin{table}
\begin{center}
\caption[]{Variable stars from Huruhata (1940) in GCVS 5.1}\label{Tab:huru}

 \begin{tabular}{lllllll}
  \hline\noalign{\smallskip}
GCVS name &  Type  & Period (GCVS 4), & Type & Period (GCVS 5.1), &Spectral type&Rem. \\
          & (GCVS 4)&days             &(GCVS 5.1)&days&(GCVS 5.1)\\
          \hline\noalign{\smallskip}
V445 Cen & E  & 14.1  &  EB:  & 28.3935    &--&\\
V446 Cen & RR & 0.52  &  RRAB &   0.5613849&--&\\
V447 Cen & RR & 0.46  &  EB   & 0.638603   &--&\\
V449 Cen & SRD& 123   &  SR   & 57.9       &G8&1\\
V452 Cen & RR & 0.406 &  RRAB &   0.7500541&--&\\
V453 Cen & SR & 118   &  SRB: &   90.6:    &--&\\
V454 Cen & RR & 0.52  &  EA:  &   0.4248050&--&\\
V466 Cen & RR & 0.573 &  RRAB &   0.3973631&--&2\\
V469 Cen & RR & 0.538 &  RRAB &   0.5812984&--&\\
V470 Cen & RR & 0.420 &  RRAB &   0.7239056&--&\\
V483 Cen & RR & 0.610 &  RRAB &   0.5713941&--&\\
\hline\noalign{\smallskip}
\multicolumn{7}{l}{\bf Remarks.}\\
\multicolumn{7}{l}{1. {\it V449 Cen.} The classification ``SRD'' was based on the G8 spectral type in the literature, but the 2MASS}\\
\multicolumn{7}{l}{infrared colors suggest a much later spectral type.}\\
\multicolumn{7}{l}{2. {\it V466 Cen}. This RRAB star has a somewhat unusual period and a very large amplitude,}\\
\multicolumn{7}{l}{$13.62-14.90^m$~$CV$.}\\
          \hline\noalign{\smallskip}
\end{tabular}
\end{center}
\end{table}

\newpage

\begin{table}
\begin{center}
\caption[]{Variable stars from McLeod and Swope (1941) in GCVS
5.1}\label{Tab:huru}

 \begin{tabular}{lllllll}
  \hline\noalign{\smallskip}
GCVS name &  Type  & Period (GCVS 4), & Type & Period (GCVS 5.1), &Spectral type&Rem. \\
          & (GCVS 4)&days             &(GCVS 5.1)&days&(GCVS 5.1)\\
          \hline\noalign{\smallskip}
V504 Cen&RCB  &-- &NL&--&pec(e)& 1\\
V517 Cen&L    &-- &SRB&178.0&--\\
V523 Cen&RR   &0.38091&RRAB&0.616281&--\\
V525 Cen&LB   &-- &SRB&108.9&--\\
V528 Cen&L    &-- &SR &147 &--\\
V531 Cen&LB   &-- &SRB&205.5&M6-7e\\
V535 Cen&RR   &0.37150& RRAB&0.5920464&--\\
V537 Cen&LB   &--&SRB&302.1&--\\
V538 Cen&RR   &0.60022&RRC&0.3747118&--\\
V541 Cen&M:   &196 &SRB& 200.8&--\\
V563 Cen&RRAB &1.07638& CWB &1.0769065&--\\
          \hline\noalign{\smallskip}
\multicolumn{7}{l}{\bf Remark.}\\
\multicolumn{7}{l}{1. {\it V504 Cen.} Definitely classified as a VY Scl cataclysmic variable by Kato and Stubbings (2003).}\\
          \hline\noalign{\smallskip}
\end{tabular}
\end{center}
\end{table}

\begin{table}
\begin{center}
\caption[]{Variable stars from Ninger-Kosybowa (1949) in GCVS
5.1}\label{Tab:nkos}
 \begin{tabular}{llllll}
  \hline\noalign{\smallskip}
GCVS name &  Type  & Period (GCVS 4), & Type & Period (GCVS 5.1), &Spectral type\\
          & (GCVS 4)&days             &(GCVS 5.1)&days&(GCVS 5.1)\\
          \hline\noalign{\smallskip}
V647 Cen  &  SR & 210& SRB &198.7&   --\\
V648 Cen  &  L: & -- &  SRB:&    59:& --\\
V649 Cen  &  L: & -- &  EW:& 0.454410:& --\\
V650 Cen  &  M  & 300:&    M&   410& Me\\
V651 Cen  &  SR & 400:&    SRB:&    57:& --\\
V652 Cen  &  SR & 86.38&   SRD& 86.9&    --\\
V653 Cen  &  IS:& -- &  EW&  0.3648759 & --\\
V654 Cen  &  IS & -- &  EW&  0.3726044 & --\\
          \hline\noalign{\smallskip}
\end{tabular}
\end{center}
\end{table}


\newpage

\begin{thebibliography}{99}

\bibitem[2009]{drake}Drake A. J., Djorgovski S. G., Mahabal A., Beshore E.,
Larson S., Graham M. J., Williams R., Christensen E., Catelan M.,
Boattini A., Gibbs A., Hill R., Kowalski R. 2009, \apj, 696, 870

\bibitem[1973]{ebisch}Ebisch K. E. 1973, \pasp, 85, 746

\bibitem[1975]{elvi}Elvius A. 1975, \aap, 44, 117

\bibitem[1930]{hofl30}Hoffleit D. 1930, Harvard Obs. Bull., No. 874, 13

\bibitem[2002]{hofl02}Hoffleit D. 2002, {\it Misfortunes as Blessings in Disguise}, American
Association of Variable Star Observers, Cambridge, MA, 176 pp.

\bibitem[1963]{hofm}Hoffmeister, C. 1963, Ver\"off. Sternwarte Sonneberg, 6, No. 1

\bibitem[1940]{huru}Huruhata M. 1940, Harvard Obs. Bull., No. 913, 14

\bibitem[2003]{kato}Kato T., Stubbings R. 2003, IBVS, No. 5426

\bibitem[2007]{kp07}Kazarovets E. V., Pastukhova E. N. 2007, Perem. Zvezdy Prilozhenie/Variable Stars Supplement, 7, No. 14

\bibitem[2008]{kp08a}Kazarovets E. V., Pastukhova E. N. 2008a, Perem. Zvezdy Prilozhenie/Variable Stars Supplement, 8, No. 24

\bibitem[2008]{kp08b}Kazarovets E. V., Pastukhova E. N. 2008b, Perem. Zvezdy Prilozhenie/Variable Stars Supplement, 8, No. 51

\bibitem[2009]{kp09}Kazarovets E. V., Pastukhova E. N. 2009, Perem. Zvezdy Prilozhenie/Variable Stars Supplement, 9, No. 32

\bibitem[2005]{kps05}Kazarovets E. V., Pastukhova E. N., Samus N. N. 2005, Perem. Zvezdy/Variable Stars, 25, No. 2

\bibitem[1998]{dnsv}Kazarovets E. V., Samus N. N., Durlevich O. V. 1998, IBVS, No. 4655

\bibitem[1985]{khol}Kholopov P. N., Samus N. N., Goranskii V. P., Gorynya N. A.,
Kireeva N. N., Kukarkina N. P., Kurochkin N. E., Medvedeva G. I.,
Perova N. B., Frolov M. S., Shugarov S. Yu., Kazarovets E. V.,
Rastorguev A. S., Karitskaya E. A., Tsvetkova T. M. et al.
1985--1995, General Catalogue of Variable Stars, 4th edition
(Nauka, Kosmosinform), Vols. I--V

\bibitem[1982]{kuk}Kukarkin B. V., Kholopov P. N., Artyukhina N. M.,
Fedorovich V. P., Frolov M. S., Goranskij V. P., Gorynya N. A.,
Karitskaya E. A., Kireeva N. N., Kukarkina N. P., Kurochkin N. E.,
Medvedeva G. I., Perova N. B., Ponomareva G. A., Samus' N. N.,
Shugarov S. Yu. 1982, New Catalogue of Suspected Variable Stars
(Moscow: Nauka)

\bibitem[1983]{maivog}Maitzen H. M., Vogt N. 1983, \aap, 123, 48

\bibitem[1941]{mcls}McLeod N. W., Swope H. H. 1941, Harvard Obs. Bull., No. 915,
29

\bibitem[1977]{moff}Moffat A. F. J. 1977, IBVS, No. 1265

\bibitem[1949]{nkos}Ninger-Kosybowa, S., 1949, Comm. from Wroclaw Observatory, No. 1

\bibitem[2006]{paspor}Paschke A., Poretti E. 2006, Open European Journal Var. Stars, No. 40

\bibitem[2007]{pas}Pastukhova E. N. 2007, Perem. Zvezdy Prilozhenie/Variable Stars Supplement, 7, No. 8

\bibitem[2000]{patea}Patterson J., Walker S., Kemp J., O'Donoghue D., Bos M., Stubbings R. 2000, \pasp, 112, 625

\bibitem[1997]{poj}Pojmanski G., 1997, AcA, 47, 467

\bibitem[2015]{sa15}Samus N. N., Antipin S. V. 2015, Highlights in Astronomy, 16,
687

\bibitem[2002]{coord}Samus' N. N., Goranskii V. P., Durlevich O. V., Zharova A. V.,
Kazarovets E. V., Pastukhova E. N., Hazen M. L., Tsvetkova T. M.
2002, Astronomy Letters, 28, 174

\bibitem[2017]{sea17}Samus' N. N., Kazarovets E. V., Durlevich O. V., Kireeva N. N.,
Pastukhova E.N. 2017, Astronomy Reports, 61, 80

\bibitem[1954]{shap}Shapley H., Allen L.B., Greenstein N., 1954, \aj, 59, 271

\bibitem[1967]{stroh}Strohmeier W. 1967, IBVS, No. 216

\bibitem[2007]{wat}Watson C. L., Henden A. A., Price A. 2007, Journal of the
AAVSO, 35, 414

\bibitem[1975]{wae}W\"alder M. 1975, Ver\"off. Remeis-Sternw.
Bamberg, 10, No. 108

\bibitem[2010]{zor}Zorotovic M., Catelan M., Smith H. A., Pritzl B. J., Aguirre
P., Angulo R. E., Aravena M., Assef R. J., Contreras C., Cortes
C., De Martini G., Escobar M. E., Gonzalez D., Jofre P., Lacerna
I., Navarro C., Palma O., Prieto G. E., Recabarren E., Trivino J.,
Vidal E. 2010, \aj, 139, 357

\end{thebibliography}
\end{document}